\documentclass[10pt]{article}
\usepackage{amsmath,amssymb,amsfonts}
\usepackage{units}
\usepackage{bbold,euler}
\usepackage{graphicx}
\usepackage{dcolumn}
\usepackage{bm,bbm,mathtools,esvect}
\usepackage{slashed}
\usepackage{esvect}
\usepackage{fullpage}
\usepackage[utf8]{inputenc}
\title{\bf KLEIN-GORDON EQUATION WITHIN THE \\ REAL HILBERT SPACE FORMALISM\\ \vspace{10mm}}

\author{ {\sf CRISTIANO ROSA}\footnote{\tt 00303944@ufrgs.br} \\
	\small \it Departamento de F\'isica \\ \\	
	{\sf SERGIO GIARDINO\footnote{\tt sergio.giardino@ufrgs.br}}\\
\small \it Departamento de Matem\'atica Pura e Aplicada \\ \\
\small \it Universidade Federal do Rio Grande do Sul (UFRGS)\\
\small \it Caixa Postal 15080, 91501-970  Porto Alegre RS \\
\small \it Brazil}
\begin{document}
\date{} 
\maketitle
\begin{abstract}
\noindent Within this article one finds the statement of the Klein-Gordon problem within the real Hilbert space formalism ($\mathbbm R$HS) in terms of complex wave functions, and in terms of quaternionic wave functions as well. The complex formulation comprises hermitian and non-hermitian cases, while the quaternionic solutions additionally set in motion self-interacting particles. The non-hermitian cases comprise non-conservative processes, while the self-interaction physically implies the increase of the effective mass of the particle, an effect that cannot be reproduced using a complex wave function. The obtained autonomous particle solutions, as well as the Klein problem agree to the previously discovered self-interacting non-relativistic particle, and thus reinforce $\mathbbm R$HS as viable and consistent way to explore open problems in quantum mechanics. Also important, the negative energy problem that plagues the usual formalism is eliminated within this approach.

\vspace{1mm}


\vspace{1mm}


\end{abstract}


\pagebreak

\hrule

\tableofcontents
\vspace{1cm}
\hrule


\section{INTRODUCTION\label{OI}}

Recently, a generalized formulation of quantum mechanics has been proposed in terms of a real Hilbert space ($\mathbbm R$HS) \cite{Giardino:2018rhs}. In fact, such a theory appeared  within the ambit of quaternionic quantum mechanics ($\mathbbm H$QM) as a way to resolve well known problems of the ill-defined classical limit of the quaternionic theory ({\em cf.} Sec. 4.4 of \cite{Adler:1995qqm}). This previous quaternionic quantum theory takes benefit of anti-hermitian Hamiltonian operators, and of a quaternionic Hilbert space. The more recent theory does not restrict the Hamiltonian operators to the Hermitian case, and  their admission of more general operators immediately confers to it a higher degree of generality. A quantum theory of quaternionic wave functions also holds for complex wave functions as a particular case,  additionally enclosing the complex quantum mechanics ($\mathbbm C$QM) within the real Hilbert space, or simply becoming the quantum mechanical theory within the $\mathbbm R$HS. However, one has to remember that this $\mathbbm R$HS theory does not correspond to a real quantum mechanics ($\mathbbm R$QM), that uses real wave functions, and employs a different formalism \cite{Renou:2021dvp,Chen:2021ril,Wu:2022vvi,Finkelstei:2022rqm,Chiribella:2022dgr,Zhu:2020iml,Fuchs:2022rih,Vedral:2023pij}.

In this article, one proposes to consider the $\mathbbm R$HS approach to the Klein-Gordon equation (KGE), using complex wave functions, as well as quaternionic wave functions, closely following the non-relativistic autonomous particle approach \cite{Giardino:2024tvp}. One  also extends the previous study of the quaternionic Klein-Gordon equation \cite{Giardino:2021lov} to non-stationary solutions, and to self-interacting solutions as well.

The historical record of attempts to generalize the KGE encompasses a varied set of ideas, including the anti-hermitian theory \cite{Adler:1995qqm,Pathak:2024rov}, quaternionic operators and complex wave functions  \cite{Edmonds:1974yq}, scalar theories \cite{DeLeo:1991mi,DeLeo:1995xt,Chanyal:2017xqf},  fermionic proposals \cite{Ulrych:2013hsq}, octonions \cite{Chanyal:2017ull},  sedenions \cite{Mironov:2018bgx,Mironov:2020lnp}, and hyperbolic hyper-complexes \cite{Ulrych:2010ac,Ulrych:2014eoa}. On the other hand, the research on solutions of the KGE within the usual complex Hilbert space remains active, and one mentions themes like
Aharonov-Bohm effect \cite{Lutfuoglu:2020wjy}, alternative quantization methods \cite{Benarab:2023gdy},
confined solutions \cite{DeVincenzo:2023jdc,Alberto:2017pkj}, conceptual understanding \cite{Bussey:2022vpi}
cosmological application \cite{Matone:2021rfj}, fermionic solutions \cite{Ferro-Hernandez:2023ymz}
Klein-Fock-Gordon equation \cite{DeVincenzo:2023jdc}, Lorentz violation \cite{Altschul:2022che}
non-linearity \cite{Raban:2022usl,Rego-Monteiro:2024ppd}, oscillators \cite{Pereira:2024cfa,Mustafa:2024cdu,Junker:2021gmv,Lutfuoglu:2020wjy,Popov:2024xox,Popov:2024wrc}, phase space \cite{Liang:2023skt}, relativistic mass \cite{Gourdain:2025foh}, super-symmetry \cite{Junker:2021gmv,Popov:2024wrc}, unification theories \cite{Hojman:2021tum}, and many other possibilities,  and thus meaning that a generalized formulation of the KGE would have a wide application range.

The $\mathbbm R$HS approach to be applied in this article has been developed within the ambit of quaternionic $\mathbbm H$QM as an attempt ot overcome the inconsistencies of the anti-hermitian theory, achieving suitable versions of the Ehrenfest theorem \cite{Giardino:2018lem}, of the Virial theorem \cite{Giardino:2019xwm,Giardino:2025bym},  and of the spectral theorem  \cite{Giardino:2018rhs}. Also applications not accessible through the anti-hermitian approach could be obtained, like the Aharonov-Bohm effect \cite{Giardino:2016xap}, the free particle \cite{Giardino:2024tvp,Giardino:2017yke,Giardino:2017pqq}, the square well \cite{Giardino:2020cee}, the Lorentz force \cite{Giardino:2019xwm,Giardino:2025bym,Giardino:2020uab}, the quantum scattering \cite{Hasan:2020ekd,Giardino:2020ztf}, and the harmonic oscillator \cite{Giardino:2021ofo}. Additionally, relativistic solutions have been obtained, including the Klein-Gordon equation \cite{Giardino:2021lov}, the Dirac equation \cite{Giardino:2021mjj}, the scalar field \cite{Giardino:2022kxk}, and the Dirac field \cite{Giardino:2022gqn}, although these relativistic solutions do not entertain the self-interacting situations entertained within this article. In summary, the aim of this article is to revisit the previous KGE solution \cite{Giardino:2021lov} using the non-relativistic autonomous particle solution \cite{Giardino:2024tvp} as a guide that will enable the introduction of non-stationary, non-hermitian, as well as self-interacting solutions.

\section{COMPLEX EQUATIONS OF MOTION\label{EOM}}

Beginning with  a relativistic particle of mass $m$, whose 	quantum motion is governed by a complex wave function $\phi=\phi(t,\,\bm x)$ of time variable $t$ and position vector $\bm x$ governed by {\it Klein-Gordon equation} (KGE),
\begin{equation}\label{kgs00}
	\left(\hbar^2\Box+m^2c^2\right)\phi=0.
\end{equation}
Of course, $c$ and $\hbar$ corresponds respectively to the speed of the light, and to the Planck's constant, while the {\it D'Alembertian operator} reads
\begin{equation}
\Box=\frac{1}{c^2}\frac{\partial^2}{\partial t^2}-\nabla^2.
\end{equation}
The physical content of this equation has to be reviewed in terms of the real Hilbert space, where  \cite{Giardino:2018rhs} to a physical observable associated with an operator $\widehat{\mathcal O}$ corresponds the {\it expectation value}
\begin{equation}\label{kgs02}
	\big\langle\mathcal O\big\rangle =\int \Big\{\psi,\,\widehat{\mathcal O}\psi\Big\}d\bm r,
\end{equation}
where one defines the {\it physical observable density} to be
\begin{equation}\label{kgs07}
\Big\{\psi,\,\widehat{\mathcal O}\psi\Big\}=\frac{1}{2}\left[\psi^\dagger \widehat{\mathcal O}\psi+\Big(\widehat{\mathcal O}\psi\Big)^\dagger\psi\right]
\end{equation}
remembering $\psi=\psi(t,\,\bm x)$ as a complex wave function, and $\psi^\dagger$ as their conjugate. One remembers (\ref{kgs07}) to be a natural inner product involving complex numbers deployed also in different contexts \cite{Giardino:2024yvt,Giardino:2025jni}, so that
\begin{equation}
\big\{z,\,w\big\}=\mathfrak{Re}[z w^\dagger],
\end{equation}
where $\mathfrak{Re}$ denotes the real part of a complex number.

One also remarks the expectation value (\ref{kgs02}) to be evaluated over the real numbers, independent of the mathematical character of the operator $\widehat{\mathcal O}$. Therefore, the Hermiticity constraint imposed to quantum operators in order to have real expectations values is removed, and the mathematical generality of the whole theory is immediately upraised. Using the contravariant {\it position four-vector} 
\begin{equation}
	x^\mu=\big(ct,\,\bm x\big),\qquad\mbox{and}\qquad \mu=\big\{0,\,1,\,2,\,3\big\},
\end{equation}
one remembers from \cite{Giardino:2021lov} the {\em continuity equation} as easily obtained from (\ref{kgs00}) to be
\begin{equation}\label{kgs01}
	\partial_\mu J^\mu=0,
\end{equation}
where the {\it probability density four-current}, defined in terms of (\ref{kgs07}) reads
\begin{equation}\label{kgs06}
 J^\mu=\Big\{\phi,\,\widehat p^\mu \phi\Big\}.
\end{equation}
Of course, $\widehat p^\mu$ is the covariant {\it linear four-momentum operator}
\begin{equation}\label{kgs05}
	\widehat p^\mu=\left(\frac{1}{c}\widehat E,\,\widehat{\bm p}\right),
\end{equation}
whose components are the energy operator, and the linear momentum operator, namely
\begin{equation}\label{kgs04}
\widehat E=i\hbar\frac{\partial}{\partial t}\qquad\mbox{and}\qquad \widehat{\bm p}=-i\hbar\bm\nabla,
\end{equation}
and (\ref{kgs05}) also admits the notation
\begin{equation}\label{kgs16}
	\widehat p^\mu=i\hbar\partial^\mu.
\end{equation}
Separating the time and space derivatives, one rewrites (\ref{kgs01}) as
\begin{equation}
	\frac{1}{c}\frac{\partial\,\rho}{\partial t}+\bm{\nabla\cdot j}=0,
\end{equation}
where the probability density four-current (\ref{kgs06}) decompose as
\begin{equation}
	 J^\mu=\Big(\rho,\,\bm{j}\Big),
\end{equation}
comprising as components the energy density $\rho$, and the linear momentum density $\bm j$, both of them generated from de definition (\ref{kgs07}) as
\begin{equation}\label{kgs03}
	\rho=\Big\{\psi,\,\widehat E\psi\Big\},\qquad\mbox{and}\qquad
	\bm j=\Big\{\psi,\,\widehat{\bm p}\psi\Big\}.
\end{equation}
This $\mathbbm R$HS formulation permits to interpret the continuity equation (\ref{kgs01}) as describing the flux of the energy-momentum density. Throughout this article, one considers the KGE (\ref{kgs00}) and the continuity equation (\ref{kgs01}) as the {\it equations of motion} of the Klein-Gordon problem.

Of course, one can generalize the physical context expressing the KGE in terms of the linear four-momentum operator (\ref{kgs05}) 
\begin{equation}\label{kgs32}
	\Big(\widehat p_\mu\widehat p^\mu -m^2c^2\Big)\phi=0,
\end{equation}
and introducing the {\it generalized linear four-momentum operator} $\widehat \pi^\mu$
\begin{equation}\label{kgs09}
	\widehat{\pi}^\mu=\widehat p^\mu-\frac{q}{c}A^\mu
\end{equation}
where $q$ is the electric charge, and $A^\mu$ corresponds to the {\it potential four-vector}
\begin{equation}
	A^\mu=\Big(\varphi,\,\bm A\Big),
\end{equation}
remembering $\varphi$ to be the scalar potential, and $\bm A$ to be the vector potential. The {\it generalized KGE}
\begin{equation}\label{kgs14}
	\Big(\widehat \pi_\mu\widehat \pi^\mu -m^2c^2\Big)\phi=0,
\end{equation}
enables the continuity equation (\ref{kgs01}) to hold as
\begin{equation}\label{kgs10}
	\partial_\mu\mathcal J^\mu=0
\end{equation}
where the density four-current (\ref{kgs06}) turns into
\begin{equation}\label{kgs11}
	\mathcal J^\mu=\Big\{\phi,\,\widehat \pi^\mu \phi\Big\},
\end{equation}
an exact generalization establishing a perfect agreement between both of the cases. Finally, one rises the generality of the model to their highest degree, requiring the generalized linear momentum (\ref{kgs09}) to admit a complex gauge potential, and therefore one defines the {\it non-hermitian linear four-momentum operator} $\widehat \Pi^\mu$ as
\begin{equation}\label{kgs15}
	\widehat{\Pi}^\mu=\widehat \pi^\mu-i\frac{q}{c}B^\mu,
\end{equation}
where $B^\mu$ is of course a real four-vector potential, and the {\it non-hermitian KGE} immediately emerges as,
\begin{equation}\label{kgs12}
	\Big(\widehat \Pi_\mu\widehat \Pi^\mu -m^2c^2\Big)\phi=0.
\end{equation} 
 One stresses the non-hermiticity of the model as a direct consequence of the non-hermiticity of the linear momentum,  and the complex linear momentum operator has already been considered in the non-relativistic case \cite{Giardino:2025bym}, where the imaginary component of the gauge potential is shown to be a natural mathematical component without immediate consequences to the expectation value. One also notices the gauge character of the imaginary components of the linear momentum as previously ascertained in \cite{Giardino:2020uab}. On the other hand, the continuity equation does not match exactly with (\ref{kgs01}) and (\ref{kgs10}), but conversely  becomes
\begin{equation}\label{kgs13}
	\Big(\partial_\mu-2\frac{q}{c}B_\mu\Big)\mathfrak J^\mu=0
\end{equation}
where the probability four-current 
\begin{equation}\label{kgs53}
	\mathfrak J^\mu=\Big\{\phi,\,\widehat \Pi^\mu \phi\Big\},
\end{equation}
fits (\ref{kgs06}) and (\ref{kgs11}). The non-hermitian character of the linear momentum operator (\ref{kgs15}) generates a non-homogeneous term on the continuity equation 
(\ref{kgs13}), denouncing the presence the non-conservative physical processes. It is essential to remind that such kind of non-homogeneous term is already known from non-relativistic $\mathbbm C$QM for complex potentials \cite{Schiff:1968qmq}, as well as to real Hilbert space QM \cite{Giardino:2018rhs}, and it is always interpreted as non-stationary phenomena, such that involving inelastic scattering, or else creation and annihilation of particles. On has to notice the dynamical character of the non-homogeneous term of (\ref{kgs13}), that is not present in the non-relativistic continuity equation \cite{Giardino:2018rhs}, whose non-homogeneous term does not depend on derivatives of the wave function. The implications of the dynamical character of the non-homogeneous term of the relativistic continuity equation is an interesting direction for future investigation. 

Summarily, in this section one defined three versions of the KGE, namely the usual (\ref{kgs00}), the generalized (\ref{kgs14},) and the non-hermitian (\ref{kgs12}), and each wave equation has its own linear momentum operator, its own probability density four-current, and its own continuity equation. Each dynamical case will be considered in the next section.

\section{COMPLEX WAVE FUNCTIONS }

Following the non-relativistic autonomous particle solution \cite{Giardino:2024tvp}, one considers the complex wave function
\begin{equation}\label{kgs08}
	\phi(t,\,\bm x)=\phi_0\exp\left[\frac{1}{\hbar}Q_\mu x^\mu\right],
\end{equation}
where $\phi_0$ is a complex amplitude, and $Q^\mu$ is a complex four-vector defined in terms of the real vectors $P^\mu$ and $K^\mu$ such as
\begin{equation}\label{kgs33}
Q^\mu=P^\mu+iK^\mu.
\end{equation}
According to the previous section, there are three cases to consider: the usual, the generalized and the non-hermitian. 

\paragraph{USUAL SOLUTION} The usual KGE wave equation (\ref{kgs00}), and the wave function (\ref{kgs08}) generate two real constraints
\begin{equation}\label{kgs35}
	K_\mu K^\mu-P_\mu P^\mu=m^2c^2\qquad\mbox{and}\qquad K_\mu P^\mu=0,
\end{equation}
as well as the continuity equation (\ref{kgs01}) produce
\begin{equation}
	K_\mu P^\mu=0.
\end{equation}
The simplest solution requires
\begin{equation}\label{kgs36}
	P^\mu=0,
\end{equation}
thus recovering the usual complex solution, and therefore the wave function (\ref{kgs08}) only admits purely oscillating physical solutions when governed by the KGE (\ref{kgs00}), recovering the usual complex Hilbert space approach, as expected.

If $P^\mu\neq 0$, one recalls that this four vector relates with non-oscillating character of the wave function. One can understand role of $P_\mu P^\mu$ in (\ref{kgs35}) as either increasing or decreasing the effective mass of the particle, and the observable effect would be the change in the amplitude associated to the particle. However, there is no physical interaction neither to damp nor to force the oscillating character of the particle in this precise case, and consequently $P^\mu=0$ is the reasonable physical choice. However, in more general situation, where interactions are present, a non-oscillating character may appear in the wave function.

\paragraph{GENERALIZED SOLUTION} In this situation, the generalized KGE (\ref{kgs14}) and the wave function (\ref{kgs08}) generate the complex constraint
\begin{equation}
	Q_\mu Q^\mu-\frac{q^2}{c^2}A_\mu A^\mu+m^2c^2+i\frac{q}{c}\Big(\hbar\partial_\mu A^\mu+2A_\mu Q^\mu\Big)=0,
\end{equation}
that separates into two real constraints
\begin{eqnarray}
\label{kgs17}&&	\left(K_\mu+\frac{q}{c}A_\mu\right)\left(K^\mu+\frac{q}{c}A^\mu\right)-P_\mu P^\mu=m^2c^2\\
\label{kgs18}&& \hbar\frac{q}{c}\partial_\mu A^\mu+2\left(K_\mu+\frac{q}{c}A_\mu\right) P^\mu=0.
\end{eqnarray}
Furthermore, the continuity equation (\ref{kgs10}) additionally determines that
\begin{equation}\label{kgs19}
	 \left(K_\mu+\frac{q}{c}A_\mu\right) P^\mu=0,
\end{equation}
and consequently (\ref{kgs18}) becomes
\begin{equation}\label{kgs26}
\partial _\mu A^\mu=0.
\end{equation}
 One stresses (\ref{kgs26}) not to be a gauge condition,  but a consistency requirement to solve (\ref{kgs17}-\ref{kgs18}), although it is formally similar to the Lorentz gauge condition. 
Moreover, there is a complete similarity between the previous case, determined by (\ref{kgs35}), and the solution composed by (\ref{kgs17}) and (\ref{kgs19}), where $P^\mu\neq 0$ can be admitted because  $A^\mu\neq 0$. Furthermore, (\ref{kgs26}) solutions assume  the form
\begin{equation}\label{kgs22}
A^\mu=A_0^\mu\exp\left[\frac{1}{\hbar} R_{0\mu} x^\mu\right],
\end{equation}
provided that $A_0^\mu$ and $R_0^\mu$ are constant real four-vectors. The simplest solution, where $R_0^\mu=0$, accepts the particular solution
\begin{equation}\label{kgs20}
	K^\mu=\alpha_0\frac{q}{c} A^\mu,\qquad A_\mu P^\mu=0,
\end{equation}
where $\alpha_0$ is a dimensionless real parameter. Therefore, (\ref{kgs17}) goes to
\begin{equation}\label{kgs21}
	P_\mu P^\mu=\big(\alpha_0-1\big)^2\left(\frac{q}{c}\right)^2A_\mu A^\mu-m^2c^2.
\end{equation}
Remembering $A^\mu$ and $m$ to be the free parameters of the system, $P^\mu$ and $K^\mu$ have been determined by (\ref{kgs19}), (\ref{kgs20}), and (\ref{kgs21}). Another solution, where $R_0^\mu\neq 0$ in (\ref{kgs22}) imposes
\begin{equation}
A_{0\mu}R_0^\mu=A_{0\mu}P^\mu=P_\mu K^\mu=0.
\end{equation}
As a particular  solution, 
\begin{equation}
 K^\mu=\alpha_0\frac{q}{c}A_0^\mu,\qquad \mbox{and}\qquad	R_0^\mu=\beta_0 P^\mu,
\end{equation}
where $\alpha_0$ and $\beta_0$ are real dimensionless parameters, recovers (\ref{kgs21}) and indicates that both of the solutions have a similar physical content. One may additionally comment the solution where
\begin{equation}
	\left(K_\mu+\frac{q}{c}A_\mu\right)\left(K^\mu+\frac{q}{c}A^\mu\right)=0
\end{equation}
leading (\ref{kgs17}) to
\begin{equation}\label{kgs23}
	P_\mu P^\mu=-m^2c^2,
\end{equation}
which is similar to the relativity energy relation, but the mass presents a flipped signal. Situations like this are explained in the usual complex theory in terms of a phantasmagorical imaginary linear momentum. However in the $\mathbbm R$HS this situation has nothing special to consider, it is only a non-propagating solution, as will be further explained in terms of the expectation values, and within the discussion of the Klein problem as well.

\paragraph{NON-HERMITIAN SOLUTION} Following the procedure adopted in the case of the generalized KGE for the non-hermitian wave equation (\ref{kgs12}) and using the wave function (\ref{kgs08}), the real constraints alike (\ref{kgs17}-\ref{kgs18}) are
\begin{eqnarray}
\label{kgs24} && \left(K_\mu+\frac{q}{c}A_\mu\right)\left(K^\mu+\frac{q}{c}A^\mu\right)-\left(P_\mu-\frac{q}{c}B_\mu\right)\left(P^\mu-\frac{q}{c}B^\mu\right)+\hbar\frac{q}{c}\partial_\mu B^\mu=m^2c^2\\
\label{kgs25} && \hbar\frac{q}{c}\partial_\mu A^\mu+2\left(K_\mu+\frac{q}{c}A_\mu\right)\left(P^\mu-\frac{q}{c}B^\mu\right)=0.
\end{eqnarray}
The continuity equation (\ref{kgs13}) generates
\begin{equation}\label{kgs27}
\left(K_\mu+\frac{q}{c}A_\mu\right)\left(P^\mu-\frac{q}{c}B^\mu\right)=0,
\end{equation} 
which implies the gauge condition (\ref{kgs26}) after their substitution in (\ref{kgs25}). It is remarkable  the energy relations  (\ref{kgs35}), (\ref{kgs17}), and (\ref{kgs24}) to have a similar structure, and the potential four vector also presents a common solution (\ref{kgs22}). Relation (\ref{kgs24}) additionally contains an association between $P^\mu$ and $B^\mu$ in the same fashion as that between $K^\mu$ and $A^\mu$, that has to be investigated in future research. A further characterization of these models can be obtained in terms of the physical observables to be entertained in the sequel.

\paragraph{EXPECTATION VALUES: USUAL CASE}

The wave function (\ref{kgs08}) and the definition of the expectation values (\ref{kgs02}) permit to determine physical observable quantities, and in the present case the energy and linear momentum operators (\ref{kgs05}) are the available possibilities to be considered.
\begin{eqnarray}
&& \big<  E\big>=-cK_0\int\!|\phi|^2 d\bm x,\\
&& \big< \bm p\big>= \bm K_1 \int\!|\phi|^2 d\bm x,
\end{eqnarray}
where
\begin{equation}\label{kgs28}
	|\phi|^2=|\phi_0|^2\exp\left[\frac{2}{\hbar}P_\mu x^\mu\right].
\end{equation}
Remembering that $P^\mu=\big(P_0,\,\bm P)$, one observes the wave function to be normalizable whenever $\bm P\neq \bm 0$ determines a decreasing space dependence. Moreover, $P_0\neq 0$ determines a non-stationary temporal dependence that may be interpreted as damped or forced physical processes. The above results are not conform to the complex Hilbert space interpretation, where the wave function (\ref{kgs08}) generates complex expectation values that cannot be explained within that approach. Let us also consider that
\begin{eqnarray}
\label{kgs29} && \Big< E^2\Big>=c^2\Big(K_0^2-P_0^2\Big) \int\!|\phi|^2 d\bm x,\\
\label{kgs30} && \Big< \|\bm p\|^2\Big>=\Big(\|\bm K\|^2-\|\bm P\|^2\Big) \int\!|\phi|^2 d\bm 	x.
\end{eqnarray}
This result, also in conformity to the non-relativistic case \cite{Giardino:2024tvp}, imply the acceptance of negative expectation values for squared energies and linear momenta, whenever either $\,K_0^2<P_0^2\,$ or $\,\|\bm K\|^2<\|\bm P\|^2\,$. As squared linear momenta are associated to kinetic energies, the above result implies the acceptance of negative energies as well. This purely quantum characteristic cannot be explained classically, mainly because the classical outcome is obtained from the square of a real quantity, while the quantum measurement comes from the definition of an inner product evaluated of the real numbers, but that is not positively defined. This result offers an solution to the old problem of negative energies of the Klein-Gordon problem that cannot be solved within the complex Hilbert space   without a field quantization scheme. In fact, there are no physical reason to reject negative energies, simply because they may be necessary in order to describe non-stationary physical processes.  The further generalized and non-hermtian cases adhere to the usual case, as follows.

\paragraph{EXPECTATION VALUES: GENERALIZED CASE}

Using the wave function (\ref{kgs08}) and the generalized linear four-momentum (\ref{kgs09}), one obtains  the energy and momentum expectation values
\begin{eqnarray}
\label{kgs31} && \big<  E\big>=-c\left(K_0+\frac{q}{c}A_0\right)\int\!|\phi|^2 d\bm x,\\
\label{kgs34} && \big< \bm p\big>= \left(\bm K_1-\frac{q}{c}\bm A\right) \int\!|\phi|^2 d\bm x.
\end{eqnarray}
Moreover, the squared energy and linear momentum give
\begin{eqnarray}
	 && \Big< E^2\Big>=c^2\left[\left(K_0+\frac{q}{c}A_0\right)^2-\big(P_0\big)^2\right] \int\!|\phi|^2 d\bm x,\\
	 && \Big< \|\bm p\|^2\Big>=\left(\left\|\bm K-\frac{q}{c}\bm A\right\|^2-\|\bm P\|^2\right) \int\!|\phi|^2 d\bm 	x.
\end{eqnarray}

The interpretation conforms the previous example, that is recovered in the $A^\mu=0$ limit, and negative energies are again allowed, as presumed.

\paragraph{EXPECTATION VALUES: NON-HERMITIAN CASE}

Using the wave function (\ref{kgs08}) and the non-hermitian linear four-momentum (\ref{kgs15}), one obtains (\ref{kgs31}-\ref{kgs34}) respectively as the energy and linear momentum expectation values. On the other hand, the squared quantities amounts to
\begin{eqnarray}
 && \Big<E^2\Big>=c^2\left[\left(K_0+\frac{q}{c}A_0\right)^2-\left(P_0-\frac{q}{c}B_0\right)^2+\hbar\frac{q}{c}\partial_tB_0\right] \int\!|\phi|^2 d\bm x,\\
 && \Big< \|\bm p\|^2\Big>=\left(\left\|\bm K+\frac{q}{c}\bm A\right\|^2-\left\|\bm P-\frac{q}{c}\bm B\right\|^2+\hbar\frac{q}{c}\bm{\nabla\cdot B}\right) \int\!|\phi|^2 d\bm 	x.
\end{eqnarray}

As supposed, the conservation of the energy
\begin{equation}
	\frac{1}{c^2}\Big<E^2\Big>=\Big< \|\bm p\|^2\Big>+m^2c^2
\end{equation}
holds on all of the three situations. In summary, the $\mathbbm R$HS formalism is capable to model the Klein-Gordon problem in a clear and consistent way, enables the comprehension of non-stationary physical processes, and permit the understanding of the existence of negative energies. Now, one considers an application to the well known Klein problem.

\section{COMPLEX KLEIN PROBLEM \label{CKP}}

The physical circumstance to be considered is the motion an autonomous particle of mass $m$ and electric charge $q$ 
scattered by a constant potential barrier. There are two situations to consider, depending whether the potential barrier is either real or complex.

\paragraph{REAL POTENTIAL} An autonomous particle suffers the action of the potential determined by the four vector
\begin{equation}\label{kgs41}
A^\mu=\left\{
\begin{array}{lll}
\big(0,\,\bm 0\big) & \mbox{if} & x<0 \\ \\
\big(V_0,\,\bm 0\big) & \mbox{if} & 0<x,
\end{array}	
\right.	
\end{equation}
where the usual KGE (\ref{kgs32}) rules the $x<0$ region of the space, while the generalized KGE (\ref{kgs14}) governs the $0<x$ region. Of course, one uses	 $x^\mu=(ct,\,x,\,0,\,0)$. The particle is fired within the $x<0$ region, where the wave function contains terms corresponding to the incident an to the reflected motion as well, whereas the $0<x$ region uniquely contains a transmitted term.  Therefore, the wave function will be
\begin{equation}\label{kgs42}
\phi(t,\,\bm x)=\left\{
	\begin{array}{lll}
		\phi_I=\exp\left[\frac{1}{\hbar}Q_\mu x^\mu\right]  +R \exp\left[\frac{1}{\hbar} Q_\mu^\dagger x^\mu\right] & \mbox{if} & x<0 \\ \\
		\phi_{II}=T\exp\left[\frac{1}{\hbar}Q'_\mu x^\mu+i\varphi_0\right] & \mbox{if} & 0<x,
	\end{array}	
	\right.	
\end{equation}
where $Q^\mu$ and $Q'^\mu$ accompany (\ref{kgs33}), $R$ and $T$ are complex amplitudes, and $\varphi_0$ is a real phase. It is clear that (\ref{kgs35}-\ref{kgs36}) holds in the $x<0$ region, and that (\ref{kgs17}-\ref{kgs26}) hold in the $0<x$ region. Moreover, at $x=0$ holds the continuity of the probability density $|\phi|$, so that
\begin{equation}\label{kgs37}
	\phi_I(t,\,\bm 0)=\phi_{II}(t,\,\bm 0)e^{-i\varphi_0}.
\end{equation}
The $\mu=0$ components of $Q^\mu$ are immediately imposed to be equal, and thus one chooses
\begin{equation}
	Q^\mu=\Big(Q^0,\,Q,\,0,\,0\Big)\qquad\mbox{and}\qquad Q'^\mu=\Big(Q^0,\,Q',\,0,\,0\Big).
\end{equation}
Consequently, the boundary condition (\ref{kgs37}), as well as an analog requirement to the space derivative gives
\begin{equation}
	R=\frac{Q-Q'}{Q+Q'},\qquad\qquad T=\frac{2Q}{Q+Q'}e^{i\varphi_0}.
\end{equation}
In order to achieve further comprehension of the system, one resorts to the continuity equation. 
Bearing in mind that the wave function $\phi_I$ comprises the motion of an incident particle, as well as the motion of a reflected particle, while $\phi_{II}$ comprises solely a transmitted moving particle, the corresponding current densities respectively read
\begin{eqnarray}
	J^\mu=\left\{
	\begin{array}{l}
	J_{inc}^\mu=\frac{i}{2}\left(Q^\mu-Q^{\dagger\mu}\right)e^{\frac{1}{\hbar}\left(Q_\nu-Q^\dagger_\nu\right)x^\nu},\\ \\
	J^\mu_{ref}=-|R|^2J_{inc}^\mu,\\ \\ J_{trans}^\mu=\frac{i}{2}\left(Q'^\mu-Q'^{\dagger\mu}\right)|T|^2e^{\frac{1}{\hbar}\left(Q_\nu-Q^\dagger_\nu\right)x^\nu}.
	\end{array}
	\right.
\end{eqnarray}
One observes the reflected and the incident current to have flipped signals, as it is presumed because of the opposite directions of the particle beams. Considering the spatial components of the beams, one defines the reflection and transmission coefficients as
\begin{equation}
	\mathcal R=-\frac{J^1_{ref}}{J^1_{inc}}\qquad\qquad \mathcal T=\frac{J^1_{trans}}{J^1_{inc}}
\end{equation}
and thus
\begin{equation}
\mathcal R=|R|^2\qquad\qquad \mathcal T=\frac{Q'-Q'^\dagger}{Q-Q^\dagger}|T|^2 \exp\left[\frac{P'_\nu-P_\nu}{\hbar}x^\nu\right].
\end{equation}	
Therefore,
\begin{equation}
\label{kgs40}\mathcal R+\mathcal T=1+2\,\frac{Q^{\dagger 2}Q'-Q^2Q'^\dagger+\left(2\exp\left[\frac{P'_\nu-P_\nu}{\hbar}x^\nu\right]-1\right)|Q|^2\big(Q'-Q'^\dagger\big)}{\big|Q+Q'\big|^2\big(Q-Q^\dagger\big)}.
\end{equation}
Of course, $P_0=P_0'=0$ because of the boundary condition, but the obtained outcome contemplates the highest degree of generality. In order to understand these results, one firstly consider the energy relations concerning the $x<0$ region of the motion, as well as the $0<x$ section of the space, respectively concerning equations (\ref{kgs35}) and (\ref{kgs17})
\begin{eqnarray}
\label{kgs38}&&	\|\bm K\|^2c^2=E^2-m^2c^4\\
\label{kgs39}&& \Big(\|\bm K'\|^2-\|\bm P'\|^2\Big)c^2=\left(E-\frac{q}{c}V_0\right)^2-m^2c^4,
\end{eqnarray}
where one used $Q^0=iE/c$, where $E$ is a real constant.
The solution corresponding to a negative right hand term of (\ref{kgs39}), as well as $\mathcal R>1,$ $\mathcal T<0$, and $\mathcal R+\mathcal T=1$ is known as the {\it Klein paradox}, and is interpreted in terms of anti-particles moving within the $0<x$ region of the space. However, the complex Hilbert space requires $P=0$ in order to have a physical motion, and introduces a phantasmagorical imaginary linear momentum.

The $\mathbbm R$HS approach, on the other hand, permits a more reasonable clearer explanation of the phenomenon, firstly because the observable expectation value of the linear momenta are always real. Moreover, the transmission coefficient is such that  $\mathcal T\propto K'/K$, and a negative signal indicates reversed particle beams in each region of the space, indicating the presence of anti-particles. Finally, the real Hilbert space formalism encompasses non-stationary phenomena, where either creation or annihilation of particles does forces non-conservative  relations to the sum $\mathcal R+\mathcal T$. The additive term of the right hand side of (\ref{kgs40}) permits the description of such kind of process, and it is also present on the non-relativistic scattering, as expected. One can now easily extend the analysis to consider the more general situation involving complex gauge potentials.

\paragraph{COMPLEX POTENTIAL } This case generalizes the gauge potential (\ref{kgs41}), such as permitting that
\begin{equation}
A^\mu\to A^\mu=\Big(V_0+V_1i,\,\bm 0\Big),\qquad \mbox{for}\qquad 0<x.
\end{equation}
This scenario accompanies the wave function, the boundary conditions, and the current densities of the previous case, and thus all the relations (\ref{kgs42}-\ref{kgs40}) of the previous case also hold. However, (\ref{kgs24}) implies (\ref{kgs39}) not to hold. Instead, it holds that
\begin{eqnarray}\label{kgs43}
	\Big(\|\bm K'\|^2-\|\bm P'\|^2\Big)c^2 -\left(\frac{q}{c}V_1\right)^2=\left(E-\frac{q}{c}V_0\right)^2-m^2c^4.
\end{eqnarray}
The fascinating point to stress is the fact that $V_1$ can take the role of $\bm P'$ in the relation energy, and thus stationary particles are admitted as solutions in case of negative right and side of (\ref{kgs24}). Such a solution is not admitted within the generalized  hermitian case, and the physical interpretation of the $V_1$ gauge potential is a exciting direction for future investigation.

\section{QUATERNIONIC EQUATIONS OF MOTION }
As already stated, the $\mathbbm R$HS approach to quantum mechanics first appeared \cite{Giardino:2018rhs} as a way to implement a consistent quaternionic quantum mechanics ($\mathbbm H$QM). Succinctly, this theory accommodates quaternionic wave functions within a $\mathbbm R$HS framework , where the  expectation value (\ref{kgs02}) is effective.  One recalls that a quaternionic wave function $\Phi$ comprehends two complex functions, namely $\Phi_0$ and $\Phi_1$, assembled as
\begin{equation}\label{kgs44}
	\Phi=\Phi_0+\Phi_1 j,
\end{equation}
where $j$ is an the quaternionic imaginary unit. The quintessential elements of quaternionic numbers  will not be outlined in this paper, they are obtainable from widespread references \cite{Ward:1997qcn,Morais:2014rqc,Ebbinghaus:1990zah}. Nevertheless, the quaternionic wave function (\ref{kgs44}) encodes an additional complex degree of freedom determined by $\Phi_1$. Moreover, the quaternionic imaginary unit $j$, and the complex imaginary unit $i$ are anti-commutative, so that $ij=-ji$, and consequently quaternions are non-commutative numbers. 

The forthcoming issue is to establish a quaternionic wave equation, and it is natural to choose (\ref{kgs12}) as the wave equation to generalize, and the essential task is  to encounter a quaternionic replacement of the non-hermitian linear momentum operator (\ref{kgs15}). In virtue of the anti-commutativity of quaternionic units, the wave function (\ref{kgs44}) is so that
\begin{equation}
	i\Phi\neq \Phi i,
\end{equation}
and consequently from the non-relativistic theory \cite{Giardino:2024tvp} one remembers that left and right linear momentum operators were established, and consequently the relativistic quaternionic theory also requires {\it left and right linear four-momentum operators}, respectively
\begin{equation}
	\widehat{p}^\mu_L\Phi=i\hbar \partial^\mu\Phi,\qquad\mbox{and}\qquad \widehat{p}^\mu_R\Phi=\hbar\partial^\mu\Phi i.
\end{equation}
Stressing the difference between the operators to be the multiplication's side of the imaginary unit $i$, one finally settles the {\it quaternionic linear four-momentum operators} to be
\begin{equation}\label{kgs46}
	\widehat{\mathscr P}_X^\mu=\widehat p^\mu_X -\frac{q}{c}\mathscr A^\mu,\qquad\mbox{where}\qquad X=\big\{L,\,R\big\},
\end{equation}
as well as the {\it quaternionic potential four-vector} 
\begin{equation}
	\mathscr A^\mu=\mathscr A_0^\mu +\mathscr A_1^\mu j.
\end{equation}
Emphasizing $\mathscr A^\mu_0$ and $\mathscr A^\mu_1$ to be complex four-vectors, one observes that  $\mathscr A_1=0$ recovers the non-hermitian linear four-momentum operator (\ref{kgs15}), as expected. Consequently, the generalization of the non-hermitian KGE (\ref{kgs12}) to the {\it quaternionic KGE} inevitably comprehends two possibilities, so that
\begin{equation}\label{kgs45}
	\Big(\widehat{\mathscr P}_{X\, \mu}\widehat{\mathscr P}_X^\mu -m^2c^2\Big)\Phi=0.
\end{equation}
The quaternionic equation (\ref{kgs45}) can be unfolded into a system of two complex equations. Using (\ref{kgs44}), one obtains
\begin{eqnarray}
\label{kgs47} && \left(\widehat\Pi_{X\mu} \widehat\Pi_X^\mu-\frac{q^2}{c^2}\mathscr A_{1\mu}\mathscr A_1^{\dagger \mu}-m^2c^2\right)\Phi_0+\widehat\Pi_{X\mu}\left(\frac{q}{c}\mathscr A_1^\mu\Phi_1^\dagger\right)-\frac{q}{c}\mathscr A_{1\mu}j\, \widehat\Pi_X^\mu\Big(\Phi_1 j \Big)=0 \\
\label{kgs48} && \widehat\Pi_{X\mu} \left[\widehat\Pi_X^\mu\Big(\Phi_1 j\Big)\right]-\left(\frac{q^2}{c^2}\mathscr A_{1\mu}\mathscr A_1^{\dagger \mu}+m^2c^2\right)\Phi_1 j-\widehat\Pi_{X\mu}\left(\frac{q}{c}\mathscr A_1^\mu\Phi_0^\dagger j\right)-\frac{q}{c}\mathscr A_{1\mu}\,j\,\widehat\Pi_X^\mu\Phi_0 =0,
\end{eqnarray}
where the definition of the operator $\widehat \Pi^\mu_X$ follows (\ref{kgs46}), so that
\begin{equation}
	\widehat \Pi_X^\mu=\widehat p_X^\mu -\frac{q}{c}\mathscr A_0^\mu.
\end{equation}
Eliminating the quaternionic imaginary unit $j$ from (\ref{kgs47}-\ref{kgs48}) requires
\begin{equation}
\label{kgs49}	\widehat p^\mu_L\big(\phi j\big)=\big(\widehat p^\mu\phi\big) j\qquad\mbox{and}\qquad \widehat p^\mu_R\big(\phi j\big)=-\big(\widehat p^\mu\phi\big) j,
\end{equation}
comprehending two case to examine. However, before looking at the equations of motion, one has to determine the continuity equation. By analogy with (\ref{kgs53}), the {\it quaternionic current density} is defined in terms of the real quantity
\begin{equation}\label{kgs58}
	\mathfrak J_X^\mu=\Big\{\widehat{\mathscr P}_X^\mu \Phi,\,\Phi\Big\},
\end{equation}
that will be proven to satisfy the {\it quaternionic continuity equation}
\begin{equation}\label{kgs52}
	\Big(\partial_\mu+\gamma^{(X)}_\mu\Big)\mathfrak J_X^\mu=0,
\end{equation}
that must recover (\ref{kgs13}) in the complex limit, and whose $\gamma^{(X)\mu}$ non-hermiticity four-vector has to be determined accordingly. Equation (\ref{kgs52}) is obtained  for $X=L$, after multiplying (\ref{kgs45}) by $-i$ on the left side, and by $\Phi^\dagger$ on the right hand side, obtaining
\begin{equation}
\partial_\mu\Big[\big(\widehat{\mathscr P}_L^\mu\Phi\big)\Phi^\dagger\Big]-\widehat{\mathscr P}_{L\mu}\Phi\partial^\mu\Phi^\dagger+i\frac{q}{c}\mathscr A_\mu\Big[\big(\widehat{\mathscr P}_L^\mu\Phi\big)\Phi^\dagger\Big]+m^2c^2i\big|\Phi\big|^2=0,
\end{equation}
and after some manipulation
\begin{equation}
\partial_\mu\Big[\big(\widehat{\mathscr P}_L^\mu\Phi\big)\Phi^\dagger\Big]+i\,\mathscr A_\mu\Big[ 2\,\mathfrak J^\mu-\Phi\big(\widehat{\mathscr P}_L^\mu\Phi\big)^\dagger \Big]-\big(\widehat p_{L\mu}\Phi\big)\partial^\mu\Phi^\dagger+\frac{q}{c}\mathscr A_\mu\Phi\partial^\mu\Phi^\dagger+m^2c^2i\big|\Phi\big|^2=0.
\end{equation}
Taking the real part of the above equation, one obtains (\ref{kgs52}) and
\begin{equation}
	\gamma^{(L)\mu}=\frac{q}{c}\Big(i\,\mathscr A^\mu-\mathscr A^{\mu\dagger} i\Big),
\end{equation}
remarking that it has been used that $iqi+q$ is a pure imaginary quaternion. The $X=R$ case
is slightly more involved, requiring the multiplication of the right hand side of (\ref{kgs45}) by $-i\Phi^\dagger$ to obtain
\begin{equation}
\partial_\mu\Big[\big(\widehat{\mathscr P}_R^\mu\Phi\big)\Phi^\dagger\Big]-\widehat{\mathscr P}_{R\mu}\Phi\partial^\mu\Phi^\dagger+i\frac{q}{c}\frac{\mathscr A_\mu}{|\Phi|^2}\Big[\big(\widehat{\mathscr P}_L^\mu\Phi\big)\Phi^\dagger\Big]\Phi i\Phi^\dagger+m^2c^2\Phi i\Phi^\dagger=0,
\end{equation}
and after some manipulation one reaches
\begin{multline}
\partial_\mu\Big[\big(\widehat{\mathscr P}_R^\mu\Phi\big)\Phi^\dagger\Big]+2\frac{q}{c}\frac{\mathscr A_\mu}{|\Phi|^2}\Phi i\Phi^\dagger \mathfrak J^\mu-\big(\widehat p_{R\mu}\Phi\big)\partial^\mu\Phi^\dagger-\frac{q^2}{c^2}\mathscr A_\mu\mathscr A^{\dagger\mu}\Phi i\Phi^\dagger +m^2c^2\Phi i\Phi^\dagger+\\
+\frac{q}{c}\frac{\mathscr A_\mu}{|\Phi|^2}\Phi i\big(\partial^\mu\Phi^\dagger\big)\Phi i\Phi^\dagger+\frac{q}{c}\mathscr A_\mu\Phi\partial^\mu\Phi^\dagger=0.
\end{multline}
The real part of the above relation generate the continuity equation, and one emphasizes the last two terms to cancel identically after some manipulation, and remembering that $|\Phi i\Phi^\dagger|=|\Phi|^2$. Therefore, (\ref{kgs52}) is obtained for
\begin{equation}\label{kgs67}
	\gamma^{(R)\mu}=\frac{q}{c}\frac{\mathscr A^\mu \Phi i\Phi^\dagger-\Phi i\Phi^\dagger \mathscr A^{\dagger\mu}}{|\Phi|^2}.
\end{equation}
Both of the considered cases recover (\ref{kgs13}) in the complex limit, when $\mathscr A^\mu$ and $\Phi$ are both complex quantities, as expected. Nevertheless, equations (\ref{kgs45}) and (\ref{kgs52}) generalize the real Hilbert space to quaternionic wave functions. The precise differences expected in quaternionic solutions will be considered the sequel.

	\section{QUATERNIONIC WAVE FUNCTIONS \label{QWF}}

In this section, one analyzes the relativistic autonomous particle in the same token as has been done in the non-relativistic case \cite{Giardino:2024tvp}, separating into two cases according to each quaternionic linear momentum operator (\ref{kgs46}).

\subsection{LEFT QUATERNIONIC KGE} Using $X=L$ in (\ref{kgs45}) and (\ref{kgs49}), equations (\ref{kgs47}-\ref{kgs48}) become
\begin{eqnarray}
\label{kgs50}	&& \left(\widehat\Pi_{\mu} \widehat\Pi^\mu-\frac{q^2}{c^2}\mathscr A_{1\mu}\mathscr A_1^{\dagger \mu}-m^2c^2\right)\Phi_0+\frac{q}{c}\left[\widehat p_\mu \mathscr A_1^\mu-\frac{q}{c}\Big(\mathscr A_{0\mu}+\mathscr A_{0\mu}^\dagger\Big)\mathscr A_1^\mu \right]\Phi_1^\dagger=0 \\
\label{kgs51}	&&  \left(\widehat\Pi_{\mu} \widehat\Pi^\mu-\frac{q^2}{c^2}\mathscr A_{1\mu}\mathscr A_1^{\dagger \mu}-m^2c^2\right)\Phi_1 -\frac{q}{c}\left[\widehat p_\mu \mathscr A_1^\mu-\frac{q}{c}\Big(\mathscr A_{0\mu}+\mathscr A_{0\mu}^\dagger\Big)\mathscr A_1^\mu \right]\Phi_0^\dagger=0.
\end{eqnarray}
One immediately notices from the above system of equations the coupling between the complex components $\Phi_0$ and $\Phi_1$ of the quaternionic wave function (\ref{kgs44}). Recalling these elements to compose the motion of a single particle, one interprets this internal coupling as a self-interacting phenomenon, a property absent in the complex solutions of the KGE, and consequently the quaternionic case cannot be reduced to the complex case because it embodies additional physical features.

A complex wave function, where $\Phi_1=0$, eliminates the self-interaction, establishing an equation for $\Phi_0$ in (\ref{kgs50}) and an equation for $\mathscr A_1^\mu$ in (\ref{kgs51}). On the other hand, the self-interaction analogously disappears if $\mathscr A_1^\mu=0$, and in this case $\Phi_0$ and $\Phi_1$ are independent non-interacting components of the quaternionic wave function. These non-interacting solutions of the quaternionic KGE have been considered in \cite{Giardino:2021lov}, and will not be considered here, where one focuses self-interacting solutions. In fact, one analyzes two solutions.

\paragraph{FIRST SOLUTION} Apparently, the simplest solution to (\ref{kgs50}-\ref{kgs51})  consists of the wave function
\begin{equation}\label{kgs62}
	\Phi=\exp\left[\frac{1}{\hbar}Q_\mu x^\mu\right]\Big(\phi_0+\phi_1j\Big).
\end{equation}
In this case, the foremost difference between the complex components $\Phi_0$ and $\Phi_1$ comprehends the complex amplitudes $\phi_0$, and $\phi_1$. This solution immediately requires that
\begin{equation}
	\widehat p_\mu \mathscr A_1^\mu-\frac{q}{c}\Big(\mathscr A_{0\mu}+\mathscr A_{0\mu}^\dagger\Big)\mathscr A_1^\mu=0.
\end{equation}
Therefore, remembering that
\begin{equation}
	\mathscr A_0^\mu=A^\mu+B^\mu \,i,
\end{equation}
the solution for $\mathscr A_1^\mu$ must be alike (\ref{kgs22}), and thus choosing
\begin{equation}
	\mathscr A_1^\mu=\mathscr A_1^{(0)\mu}\exp\left[\frac{1}{\hbar}\frac{q}{c}H_{0\mu}x^\mu\right],
\end{equation}
where $\mathscr A_1^{(0)\mu}$ and $H_0^\mu$ are constant four vectors, one arrives at the constraint
\begin{equation}\label{kgs61}
	\Big(H_{0\mu}-2A_\mu\Big)\mathscr A_1^{(0)\mu}=0.
\end{equation}
On the other hand, both of the complex components of the quaternionic wave function, respectively $\Phi_0$ and $\Phi_1$, satisfy the constraint generated within the equations of motion (\ref{kgs50}-\ref{kgs51}), namely
\begin{multline}\label{kgs63}
\left(K_\mu+\frac{q}{c}A_\mu\right)\left(K^\mu+\frac{q}{c}A^\mu\right)-\left(P_\mu-\frac{q}{c}B_\mu\right)\left(P^\mu-\frac{q}{c}B^\mu\right)+\hbar\frac{q}{c}\partial_\mu B^\mu-\frac{q^2}{c^2}\mathscr A_{1\mu}\mathscr A_1^{\dagger \mu}-m^2c^2	=0.
\end{multline}
Meanwhile, one also has to consider from (\ref{kgs58}) the density four vector
\begin{equation}\label{kgs68}
\mathfrak J^\mu =-\Big(K^\mu+A^\mu\Big)|\Phi|^2,
\end{equation}
that within the continuity equation (\ref{kgs52}) recovers the equations of the non-hermitian case, respectively (\ref{kgs25}) and (\ref{kgs27}). Therefore, the (\ref{kgs22}) solution for the $A^\mu$ potential four vector is also common with the non-hermitian solution.  

In summary, choosing constant components of the potential four vector satisfying constraints (\ref{kgs27}) and (\ref{kgs61}), one concludes that the difference between this quaternionic solution, and the non-hermitian solution is the presence of the $\mathscr A_{1\mu}\mathscr A_1^{\dagger \mu}$ term in (\ref{kgs63}), that can be interpreted as a correction of the mass of the particle to an effective value suited to the motion. One observes that even massless solutions are possible, but they would exhibit an effective mass generated by this purely quaternionic term. The investigation whether such a field may be a generator of mass within concrete processes is a fascinating direction of future research.

As a final comment, energy and linear momentum expectation values can be obtained from
\begin{equation}\label{kgs64}
	\big< p^\mu\big>=\int \mathfrak J^\mu dx,
\end{equation}
using (\ref{kgs68}), and the complex results (\ref{kgs31}-\ref{kgs34}) are recovered whenever $\phi_1=0$. Particularly, energy and linear momentum expectation values reproduce the non-hermitian values (\ref{kgs31}-\ref{kgs34}), and a further calculation gives
\begin{equation}\label{kgs66}
	\Big<\mathscr P_{L\mu}\mathscr P_L^\mu\Big>=\big< m^2 c^2\big>,
\end{equation}
thus recovering the constraint (\ref{kgs63}) after eliminating the common exponential dependence. Conclusively, the $\mathscr A_{1\mu}\mathscr A_1^{\dagger \mu}$ term in (\ref{kgs63}) is the single difference to the previous cases. As said, the physical consequences may be interesting and should be investigated in future research. A different situation will be observed in the next solution.

\paragraph{SECOND SOLUTION}
Recalling the non-relativistic autonomous particle \cite{Giardino:2024tvp} as the prototype solution of the KGE, one proposes the wave function 
\begin{equation}\label{kgs54}
	\Phi=\Big(\phi_0+\phi_1j\Big)\exp\left[\frac{1}{\hbar}Q_\mu x^\mu\right].
\end{equation}
Compared with (\ref{kgs62}), the location of the exponential function on the right hand side makes the functional dependence of the complex component $\Phi_1$ of the wave function to be based on $Q^{\dagger \mu}$. Therefore, using (\ref{kgs54}) in (\ref{kgs50}-\ref{kgs51}) generates a system of equations for the constants $\phi_0$ and $\phi_1$ that can be written in the matrix form
\begin{equation}\label{kgs57}
	M\bm\phi=0,
\end{equation}
where 
\begin{equation}\label{kgs55}
	M=\left[
	\begin{array}{cc}
		M_{11} & M_{12} \\
		M_{21} & M_{22}
	\end{array}
	\right],\qquad 
	\bm \phi=\left[
	\begin{array}{c}
    \phi_0 \\
    \phi_1^\dagger
	\end{array}
	\right].
\end{equation}
One thus obtains
\begin{multline}
	M_{11}=\left(K_\mu+\frac{q}{c}A_\mu\right)\left(K^\mu+\frac{q}{c}A^\mu\right)-\left(P_\mu-\frac{q}{c}B_\mu\right)\left(P^\mu-\frac{q}{c}B^\mu\right)+\hbar\frac{q}{c}\partial_\mu B^\mu-\frac{q^2}{c^2}\mathscr A_{1\mu}\mathscr A_1^{\dagger \mu}-m^2c^2	-\\
	-i\Bigg[\hbar\frac{q}{c}\partial_\mu A^\mu+2\left(K_\mu+\frac{q}{c}A_\mu\right)\left(P^\mu-\frac{q}{c}B^\mu\right)\Bigg]
\end{multline}
\begin{multline}
	M_{22}=\left(K_\mu-\frac{q}{c}A_\mu\right)\left(K^\mu-\frac{q}{c}A^\mu\right)-\left(P_\mu-\frac{q}{c}B_\mu\right)\left(P^\mu-\frac{q}{c}B^\mu\right)+\hbar\frac{q}{c}\partial_\mu B^\mu-\frac{q^2}{c^2}\mathscr A_{1\mu}\mathscr A_1^{\dagger \mu}-m^2c^2	+\\
	+i\Bigg[\hbar\frac{q}{c}\partial_\mu A^\mu-2\left(K_\mu-\frac{q}{c}A_\mu\right)\left(P^\mu-\frac{q}{c}B^\mu\right)\Bigg]
\end{multline}
\begin{equation}
M_{12}=\frac{q}{c}\left(\widehat p_\mu \mathscr A_1^\mu-2\frac{q}{c}A_\mu\mathscr A_1^\mu\right),\qquad M_{21}=-M_{12}^\dagger.
\end{equation}
The solution  requires the determinant of $M$ to be zero. However, before  entertaining this calculation, one directs the attention to the probability density current (\ref{kgs58})
\begin{equation}\label{kgs65}
\mathfrak J^\mu=-\Bigg[ K^\mu\Big(|\phi_0|^2-|\phi_1|^2\Big)+\frac{q}{c}A^\mu\Big(|\phi_0|^2+|\phi_1|^2\Big)\Bigg]\exp\left[\frac{2}{\hbar}P_\mu x^\mu\right],
\end{equation}
that within the continuity equation (\ref{kgs52}) produces the constraint
\begin{equation}\label{kgs56}
\Bigg[|\phi_0|^2\left(K_\mu+\frac{q}{c}A_\mu\right)-|\phi_1|^2\left(K_\mu-\frac{q}{c}A_\mu\right)\Bigg]\left(P^\mu-\frac{q}{c}B^\mu\right)=0,
\end{equation}
that of course recovers (\ref{kgs27}) if $\phi_1\to 0$, as required. A more readable perspective can be obtained rewriting (\ref{kgs55}) as
\begin{equation}\label{kgs72}
	M=\left[
	\begin{array}{cc}
		F+G^\dagger & H \\
		-H^\dagger & F^\dagger-G
	\end{array}
	\right],
\end{equation}
where $H=M_{12}$, 
\begin{multline}
F=K_\mu K^\mu +\frac{q^2}{c^2}A_\mu A^\mu- \left(P_\mu-\frac{q}{c}B_\mu\right)\left(P^\mu-\frac{q}{c}B^\mu\right)+\hbar\frac{q}{c}\partial_\mu B^\mu-\frac{q^2}{c^2}\mathscr A_{1\mu}\mathscr A_1^{\dagger \mu}-m^2c^2	-\\
	-i\Bigg[\hbar\frac{q}{c}\partial_\mu A^\mu+2\frac{q}{c}A_\mu\left(P^\mu-\frac{q}{c}B^\mu\right)\Bigg],
\end{multline}
and
\begin{equation}
	G=2\left[\frac{q}{c}A_\mu +i\left(P_\mu-\frac{q}{c}B_\mu\right)\right]K^\mu.
\end{equation}
The simplest solution seems to be
\begin{equation}
F=0,\qquad\mbox{and}\qquad G=H,
\end{equation}
correspondingly accepting constant $A^\mu$ and $B^\mu$, so that
\begin{equation}
	K^\mu=\alpha_0\frac{q}{c}A^\mu,\qquad P^\mu=\alpha_0\frac{q}{c}B^\mu,\qquad\mbox{and}\qquad A_\mu B^\mu=0,
\end{equation}
thus fulfilling the (\ref{kgs56}) constraint for $\alpha_0$ and $\beta_0$ real constants, and thus determining the solution. However, the physical interpretation presents interesting aspects. First of all, one observes that the signal of the energy and linear momentum expectation values obtained from (\ref{kgs64}) using the density four vector (\ref{kgs65}) depends on the relation between the amplitudes $\phi_0$ and $\phi_1$, demonstrating the self-interaction to have measurable effects. Moreover, 
the conservation of the energy relation (\ref{kgs66}) gives
\begin{multline}
	\left(K_\mu+\frac{q}{c}A_\mu\right)\left(K^\mu+\frac{q}{c}A^\mu\right)-\left(P_\mu-\frac{q}{c}B_\mu\right)\left(P^\mu-\frac{q}{c}B^\mu\right)+\hbar\frac{q}{c}\partial_\mu B^\mu+
	\frac{4|\phi_1|^2}{|\phi_0|^2+|\phi_1|^2}\frac{q}{c}K_\mu A^\mu-\\
	-\frac{q^2}{c^2}\mathscr A_{1\mu}\mathscr A^{\dagger\mu}_1=m^2c^2
\end{multline}
recovering the non-hermitian energy conservation relation (\ref{kgs63}) in the limit $\phi_1\to 0$. This energy relation can be considered very interesting because it presents a further mechanism of mass generation, now related to the self-interaction term  on $\phi_1$. 

In summary, one observes within the two quaternionic solutions serious differences to the purely complex case, where mass generation mechanisms may depend on the potential four vector, and on the quaternionic self-interaction. These are novel features that cannot be obtained in terms of the purely complex cases, and thus justifying to keep the attention to the research of quaternionic solutions.

\subsection{RIGHT QUATERNIONIC KGE} Further quaternionic solutions are obtained after setting  $X=R$ in the linear momentum operator (\ref{kgs46}), and thus the KGE (\ref{kgs45}) furnishes 
a system of equations from (\ref{kgs47}-\ref{kgs48}), so that
\begin{eqnarray}
	\label{kgs59}&& \left(\widehat\Pi_{\mu} \widehat\Pi^\mu-\frac{q^2}{c^2}\mathscr A_{1\mu}\mathscr A_1^{\dagger \mu}-m^2c^2\right)\Phi_0+\frac{q}{c}\left[\widehat p_\mu \mathscr A_1^\mu +2\mathscr A_{1\mu}\widehat p^\mu-\frac{q}{c}\Big(\mathscr A_{0\mu}+\mathscr A_{0\mu}^\dagger\Big)\mathscr A_1^\mu \right]\Phi_1^\dagger=0 \\
	\label{kgs60}&&  \left(\widehat\varpi_{\mu} \widehat\varpi^\mu-\frac{q^2}{c^2}\mathscr A_{1\mu}\mathscr A_1^{\dagger \mu}-m^2c^2\right)\Phi_1 +\frac{q}{c}\left[\widehat p_\mu \mathscr A_1^\mu+2\mathscr A_{1\mu}\widehat p^\mu+\frac{q}{c}\Big(\mathscr A_{0\mu}+\mathscr A_{0\mu}^\dagger\Big)\mathscr A_1^\mu \right]\Phi_0^\dagger=0,
\end{eqnarray}
where it holds the definition
\begin{equation}
	\widehat\varpi^\mu=\widehat p^\mu+\frac{q}{c}\mathscr A^\mu_0.
\end{equation}
As in the left case, there are two possible wave functions, namely (\ref{kgs62}) and (\ref{kgs54}), and one must considers them separately again. 

\paragraph{FIRST SOLUTION} Wave function (\ref{kgs62}) in   (\ref{kgs59}-\ref{kgs60}) imposes zero coefficients, because $\Phi_0$ depends on $Q^\mu$, and $\Phi_1^\dagger$ depends on $Q^{\mu\dagger}$. A mathematically more general solution would require a functional identity between $\Phi_0$ and $\Phi_1$ attainable in terms of a real $Q^\mu$. However, this is a non-propagating case, and will not be considered. Therefore, both of the complex components $\Phi_0$ and $\Phi_1$ must generate an identical energy relation
\begin{eqnarray}\label{kgs69}
	-Q_\mu Q^\mu +\frac{q^2}{c^2}\mathscr A_{0\mu}\mathscr A_0^\mu-\frac{q^2}{c^2}\mathscr A_{1\mu} \mathscr A_1^{\dagger\mu} -m^2c^2=0,
\end{eqnarray}
as well as the constraint
\begin{equation}
\hbar\partial_\mu\mathscr A_0^\mu+2\mathscr A_{0\mu} Q^\mu=0.
\end{equation}
Finally, the further coefficients generate 
\begin{equation}
 \hbar \partial\mathscr A_{1\mu}^\mu+2\mathscr A_{1\mu}Q^{\dagger \mu}-2i\frac{q}{c} A_{\mu}\mathscr A_1^\mu=0.
\end{equation}
The solutions involve gauge four vectors alike (\ref{kgs22}), and the particular cases can be easily considered. Furthermore, the density four vector reproduces (\ref{kgs68}), whereas the non-homogeneity term (\ref{kgs67}) comprises
\begin{equation}\label{kgs70}
	\gamma^{(R)\mu}=\frac{q}{c}\frac{1}{|\Phi|^2}\left[-2 B^\mu\Big(|\Phi_0|^2-|\Phi_1|^2\Big)+4i\Big(\mathscr A_1^{\dagger\mu}\Phi_0\Phi_1-\mathscr A_1^\mu\Phi_0^\dagger\Phi_1^\dagger\Big)\right],
\end{equation}
and consequently  the continuity equation (\ref{kgs52}) generates the constraint
\begin{multline}\label{kgs71}
	\left(K_\mu+\frac{q}{c}A_\mu\right)\Bigg[\left(P^\mu-\frac{q}{c}B^\mu\right)|\phi_0|^2+\left(P^\mu+\frac{q}{c}B^\mu\right)|\phi_1|^2+\\+2i\frac{q}{c}\Bigg(\mathscr A_1^{\dagger\mu}\phi_0\phi_1 \exp\left[\frac{2i}{\hbar}K_\mu x^\mu\right]-\mathscr A_1^\mu\phi_0^\dagger\phi_1^\dagger \exp\left[-\frac{2i}{\hbar}K_\mu x^\mu\right]\Bigg)\Bigg]=0.
\end{multline}
A simple solution may require
\begin{equation}\label{kgs75}
	\mathscr A_1^\mu=\mathscr A_1^{(0)\mu}\phi_0\phi_1e^{2iK_\mu x^\mu},
\end{equation}
where $\mathscr A_1^{(0)\mu}$ is a constant complex four vector. A constraint  between $\phi_0$ and $\phi_1$ is obtained in (\ref{kgs71}) if a real $\mathscr A_1^{(0)\mu}$ is chosen, but this solution seems not to deserve further attention. Their energy relation (\ref{kgs69}) does not contain interesting physical cases to explore, despite the fact that it is important to present this solution in order to exhaust the mathematical possibilities.

\paragraph{SECOND SOLUTION} Using the (\ref{kgs54}) wave function, equations (\ref{kgs57}-\ref{kgs55}) are rewritten in terms of a novel matrix $N$ to be obtained from (\ref{kgs59}-\ref{kgs60}), so that
\begin{equation}
	N\bm\phi=0,
\end{equation}
and one obtains
\begin{equation}
	N_{11}=M_{11},
\end{equation}
\begin{multline}
	N_{22}=\left(K_\mu+\frac{q}{c}A_\mu\right)\left(K^\mu+\frac{q}{c}A^\mu\right)-\left(P_\mu+\frac{q}{c}B_\mu\right)\left(P^\mu+\frac{q}{c}B^\mu\right)-\hbar\frac{q}{c}\partial_\mu B^\mu-\frac{q^2}{c^2}\mathscr A_{1\mu}\mathscr A_1^{\dagger \mu}-m^2c^2	+ \\
	+i\Bigg[\hbar\frac{q}{c}\partial_\mu A^\mu+2\left(K_\mu+\frac{q}{c}A_\mu\right)\left(P^\mu+\frac{q}{c}B^\mu\right)\Bigg],
\end{multline}
\begin{equation}
	N_{12}=i\hbar\partial_\mu\mathscr A_1^\mu+2i\mathscr A_{1\mu}Q^\mu-2\frac{q}{c} A_\mu\mathscr A_1^\mu\qquad\mbox{and}\qquad N_{21}=N_{12}^\dagger.
\end{equation}
The second equation of motion, that is the continuity equation (\ref{kgs52}), encompasses similarities to the previous case, where the probability density four vector is equal to (\ref{kgs68}), and continuity equations generates the constraint
\begin{multline}
	\left(K_\mu+\frac{q}{c}A_\mu\right)\Bigg[\left(P^\mu-\frac{q}{c}B^\mu\right)|\phi_0|^2+\left(P^\mu+\frac{q}{c}B^\mu\right)|\phi_1|^2+2i\frac{q}{c}\Bigg(\mathscr A_1^{\dagger\mu}\phi_0\phi_1-\mathscr A_1^\mu\phi_0^\dagger\phi_1^\dagger \Bigg)\Bigg]=0,
\end{multline}
that can also be simplified using a choice like (\ref{kgs75}), but without the $K_\mu x^\mu$ functional dependence found in (\ref{kgs71}). Conform to (\ref{kgs72}), one adopts a matrix procedure to solve (\ref{kgs59}-\ref{kgs60}), so that
\begin{equation}
	N=\left[
	\begin{array}{cc}
		U+V^\dagger & W \\
		-W^\dagger & U^\dagger-V
	\end{array}
	\right],
\end{equation}
where $W=N_{12}$, 
\begin{multline}
	U=\left(K_\mu+\frac{q}{c}A_\mu\right)\left(K^\mu+\frac{q}{c}A^\mu\right)-P_\mu P^\mu -\frac{q^2}{c^2}B_\mu B^\mu-\frac{q^2}{c^2}\mathscr A_{1\mu}\mathscr A_1^{\dagger \mu}-m^2c^2	-\\
	-i\Bigg[\hbar\frac{q}{c}\partial_\mu A^\mu+2P_\mu\left(K^\mu+\frac{q}{c}A^\mu\right)\Bigg],
\end{multline}
and
\begin{equation}
	V=\frac{q}{c}\Bigg[\hbar\partial_\mu B^\mu+2\left[P_\mu +i\left(K_\mu+\frac{q}{c}A_\mu\right)\right]B^\mu\Bigg].
\end{equation}
The simplest solution seems to be
\begin{equation}
	U=0,\qquad\mbox{and}\qquad V=W.
\end{equation}
The solution follows the previous case, but more interestingly, the energy relation gives
\begin{multline}
	\left(K_\mu+\frac{q}{c}A_\mu\right)\left(K^\mu+\frac{q}{c}A^\mu\right)-P_\mu P^\mu -\frac{q^2}{c^2}B_\mu B^\mu+\hbar\frac{q}{c}\partial_\mu B^\mu-\frac{q^2}{c^2}\mathscr A_{1\mu}\mathscr A^{\dagger\mu}_1-\\
	-\frac{2}{|\phi_0|^2+|\phi_1|^2}\frac{q}{c}\Bigg[\hbar|\phi_1|^2\partial_\mu B^\mu+i\partial_\mu\Big(\phi_0\phi_1\mathscr A_1^{\dagger\mu}-\phi_0^\dagger\phi_1^\dagger\mathscr A_1^{\mu}\Big)\Bigg]=m^2c^2.
\end{multline}
The  physical interpretation is somewhat analogous to the previous solution, where a correction of the mass term appears as $\mathscr A_{1\mu}\mathscr A^{\dagger\mu}_1$. However, there  is $\phi_1$-dependent term that can be understood as a self-interaction correction to the energy relation. The physical processes associated to such a term are of course to be considered in future investigation.

\section{QUATERNIONIC KLEIN PROBLEM }

In this final section, one considers the potential alike (\ref{kgs41}), but involving quaternionic quantities,  so that
\begin{equation}
	A^\mu\to A^\mu=\Big(\mathscr A_0+\mathscr A_1 j,\,\bm 0\Big),\qquad \mbox{for}\qquad 0<x,
\end{equation}
where $\mathscr A_0$ and $\mathscr A_1$ are  complex constants. This kind of problem presents an apparent inconvenience remembering the solution to involve a wave function comparable to (\ref{kgs42}), where the $x<0$ region contains a complex solution, and a quaternionic function is comprehend within $0<x$. The boundary condition of continuous wave function that is usually found in $\mathbbm C$QM  imposes the quaternionic solution to be complex, and consequently only $\mathscr A_1=0$ solutions would satisfy such a strict condition. 

However, the physically motivated  continuity of the probability density condition conforming to (\ref{kgs37})  takes account of the situation. In the complex case, the phase difference between the two regions of the potential is a complex unitary number, and consequently an unitary quaternionic phase difference will be required. Emphasizing that a quaternionic function will replace $\phi_{II}$ in (\ref{kgs42}), the correct boundary condition in terms of an unitary quaternionic phase $\mathcal U$ will be either
\begin{equation}\label{kgs73}
	\phi_I(t,\,\bm 0)=\phi_{II}(t,\,\bm 0)\, \mathcal U,
\end{equation}
if $\phi_{II}$ conforms (\ref{kgs62}), or
\begin{equation}\label{kgs74}
	\phi_I(t,\,\bm 0)=\mathcal U\,\phi_{II}(t,\,\bm 0),
\end{equation}
if $\phi_{II}$ conforms to the second solution (\ref{kgs54}). 

These conditions turn complex both of the sides of (\ref{kgs73}-\ref{kgs74}), and hence these boundary condition can make sense. After doing this, all of the calculations developed in Section \ref{CKP} will be repeated, and the differences comprise mainly the four vectors $Q^\mu$ and $Q^{\prime\mu}$. On the other hand, energy relations of the quaternionic cases will be similar to the complex energy relation (\ref{kgs43}) of the complex case. However, the quaternionic decisive term $\mathscr A_{1\mu}\mathscr A_1^{\dagger\mu}$  has a flipped signal when compared to the complex contribution of the complex component $V_1$ of (\ref{kgs43}). Therefore, the $\mu=0$ contribution to the quaternionic term necessarily increases the effective mass of the particle, while the complex component has an opposite effect, as can be clearly seen in (\ref{kgs69}). 

Indisputably, the purely complex and the purely quaternionic components interfere oppositely within the Klein problem, and hence  attesting the non reducibility of the quaternionic case to a complex solution, and it also confirms the effect of increasing the number of degrees of freedom through quaternionic formalism as way to generalize quantum mechanics. The investigation of the precise physical interactions that can be described in terms of quaternionic wave functions throughout the $\mathbbm R$HS approach is an exciting direction of future research.

\section{CONCLUSION\label{OCO}}

In this article, one considered the Klein-Gordon problem within the real Hilbert space formalism ($\mathbbm R$HS), and the problem has been addressed with respect to complex wave functions, and to quaternionic wave functions as well. The $\mathbbm R$HS formalism has been proven to completely solve the Klein-Gordon problem in situations where the usual complex Hilbert space formalism fails. First of all, it admits the existence of negative energies, but  it also solves the Klein problem without the resort to imaginary linear momentum, it enables a consistent generalization of the linear momentum operator, and it permits the approach to non-stationary physical situations. Moreover, the $\mathbbm R$HS approach is also well suited for quaternionic wave functions, thus enabling additional degrees of freedom whose physical expression is not equivalent to that generated in terms of complex wave functions. 

Summarily, the results contained in this paper confirm the robustness of the generalization of quantum mechanics obtained in terms of the $\mathbbm R$HS program. This robustness already determined in non-relativistic cases was extended successfully to the relativistic case of the Klein-Gordon equation. The future directions of research are various, and include the investigation of concrete physical models to be described in terms of the $\mathbbm R$HS formalism, as well as the extension of the same formalism to further theoretical situations involving the Dirac equation, field quantization,  Lorentz and CPT invariance, and many other possibilities as well.

\begin{footnotesize}
\paragraph{Funding} SG gratefully thanks for the financial support by Fapergs under the grant 23/2551-0000935-8 within Edital 14/2022, and CR thanks for the scientific initiation grant by PIBIC-CNPq.

\paragraph{Data availability statement} The author declares that data sharing is not applicable to this article as no data sets were generated or analyzed during the current study.

\paragraph{Declaration of interest statement} The author declares that he has no known competing financial interests or personal relationships that
could have appeared to influence the work reported in this paper.
\end{footnotesize}
%
%
%
%
\begin{footnotesize}

\end{footnotesize}

\end{document}